\documentclass{article} % For LaTeX2e
\def\eqref#1{equation~\ref{#1}}
\def\1{\bm{1}}

\DeclareMathAlphabet{\mathsfit}{\encodingdefault}{\sfdefault}{m}{sl}
\SetMathAlphabet{\mathsfit}{bold}{\encodingdefault}{\sfdefault}{bx}{n}

\usepackage[preprint]{OneColumnPreprint}

\usepackage[utf8]{inputenc} % allow utf-8 input
\usepackage[T1]{fontenc}    % use 8-bit T1 fonts
\usepackage{hyperref}       % hyperlinks
\usepackage{url}            % simple URL typesetting
\usepackage{booktabs}       % professional-quality tables
\usepackage{amsfonts}       % blackboard math symbols
\usepackage{nicefrac}       % compact symbols for 1/2, etc.
\usepackage{microtype}      % microtypography
\usepackage{xcolor}         % colors

\usepackage{multirow}
\usepackage{wrapfig}

\usepackage{graphics}
\usepackage{graphicx}
\usepackage{amsmath}
\usepackage{wrapfig}
\usepackage[subfigure]{tocloft}
\usepackage{subcaption}
\usepackage{amsfonts,bm,amssymb}
\usepackage{natbib} 

\usepackage{macros}
\usepackage{algorithm}

\usepackage{algpseudocode}
\usepackage{amsmath}
\usepackage{float}
\usepackage{multicol}
\usepackage{setspace}

\usepackage{tikz}
\usetikzlibrary{shapes.geometric, arrows}
 \usepackage{tabularray}
 \usepackage{multirow}

\usepackage[inline]{enumitem}

\renewcommand{\cite}[1]{\citep{#1}}

\newcommand{\US}{U\d{s}\'{a}s}
\newcommand{\biul}[1]{\noindent\underline{\textbf{\textit{#1}}}}

\newcommand{\salt}{\texttt{\textit{Salient Store }}}
\title{Salient Store:  Enabling Smart Storage for Continuous Learning Edge Servers}

\author{%
Cyan Subhra Mishra, Deeksha Chaudhary, Jack Sampson, Mahmut Taylan Knademir, Chita Das\\
\texttt{The Pennsylvania State University}
\texttt{\{cyan, dmc6955, jms1257, mtk2, cxd12\}@psu.edu}
}

\begin{document}
\maketitle
%%%%%%%%%%%%%%%%%%%%%%%%%%%%%%%%%%%%%%%%%%%%%%%%%%%
%%%%%% -- PAPER CONTENT STARTS-- %%%%%%%%

\begin{abstract}
%%%%%

As continuous learning based video analytics continue to evolve, the role of efficient edge servers in efficiently managing vast and dynamic datasets is becoming increasingly crucial. Unlike their compute architecture, storage and archival system for these edge servers has often been under-emphasized. This is unfortunate as they contribute significantly to the data management and data movement, especially in a emerging complute landscape where date storage and data protection has become one of the key concerns. To mitigate this, we propose \salt that specifically focuses on the integration of Computational Storage Devices (CSDs)  into edge servers to enhance data processing and management, particularly in continuous learning scenarios, prevalent in fields such as autonomous driving and urban mobility. Our research, gos beyond the compute domain, and identifies the gaps in current storage system designs. We proposes a framework that aligns more closely with the growing data demands. We present a detailed analysis of data movement challenges within the archival workflows and demonstrate how the strategic integration of CSDs can significantly optimize data compression, encryption, as well as other data management tasks, to improve overall system performance. By leveraging the parallel processing capabilities of FPGAs and the high internal bandwidth of SSDs, \salt reduces the communication latency and data volume by $\approx6.2\times$ and $\approx6.1\times$, respectively. This paper provides a comprehensive overview of the potential of CSDs to revolutionize storage, making them not just data repositories but active participants in the computational process.

%%%%%

\end{abstract}

%%%%%% -- MAIN CONTENT STARTS-- %%%%%%%%
%-------------------------------------------------------------------------------
%\vspace{-4pt}
\section{Introduction}
\label{sec:introduction}
Video analytics, powered by deep neural networks (DNNs) has become the key component of multiple applications including but not limited to autonomous driving~\cite{autoDriving, autoDriving01, autoDriving02}, urban mobility~\cite{Intelurbanmobility, googleUrbanMobility}, surveillance and monitoring~\cite{surveillance01, surveillance02, surveillance03}, video streaming and conferencing~\cite{vidStream01, vidStream02, vidStream03}, telemedicine~\cite{telemedicine}, and tourism~\cite{tourism01, tourism02, tourism03}. While some of these applications rely on collecting the video data and processing them offline, many need real-time analytics for the seamless integration, operation and effectiveness of the task at hand~\cite{realtimeVideo, realtimeVideo01, realtimeVideo02}. Moreover, depending on the deployment, scenario and requirements, some of these applications also demand learning to keep up with the data drift~\cite{ekya, usas, dacapo, icarl}. However, regulations, resource limitations and privacy concerns often mandate these applications (both learning and inference) to be performed at the edge~\cite{ekya, usas}. For example, many of the European cities restrict the traffic video data to be streamed to the cloud~\cite{sweden-data, azure-data, ekya}, which enforces performing video analytics and learning tasks related to urban mobility at the edge. This has lead to a significant development in the direction of enabling video analytics and learning with edge servers.

Although efficient algorithms, compute orchestration and hardware have addressed the analytics part, the scaling of such  a system becomes a problem primarily due to high energy consumption. Recent works try to solve this problem by augmenting these continuous learning edge servers with application-specific hardware targeted for intermittent computing which could run using solar power. However, all these works focus on the analytics part while overlooking one critical aspect: \emph{\textbf{what happens to all those video data after the analytics?}} 

\biul{Data Archival:} The answer is straightforward, especially for mission-critical public records like urban mobility and surveillance data: these need to be archived in a local storage to avoid undermining the benefits of edge computation, i.e. minimizing communication and preserving privacy. However, managing such data requires substantial storage infrastructure. For instance, storing a full day's worth of 1080p video at 60fps requires ($1920 \times 1080 \times 3 \text{ pixels } \times 4 \text{ bytes per pixel } \times 60 \text{ frames per second } \times 3600 \times 24) 117.32$ TiB of raw video data, which compresses to approximately $60$ GiB to $400$ GiB of encoded data per day. Including redundancy, this requires an additional $33\%$ to $100\%$ more storage capacity. Furthermore, given the plug-and-play nature of storage media like HDDs and SSDs, securing this data becomes even more critical.

\begin{figure}[]
\centering 
\includegraphics[width=0.7\linewidth]{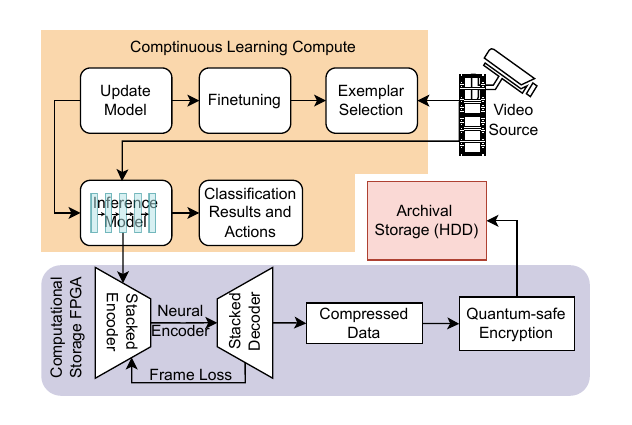}
\caption{Data flow pipeline of continuous learning edge servers with storage and data archival pipeline. The Shown storage pipeline is the preliminary focus of \salt.}
\label{fig:saltpipe}
\end{figure}

The typical \textit{archival process} involves three key phases: compression, encryption, and redundancy. Fig.~\ref{fig:saltpipe} illustrates the data flow in an edge server, where video data are first encoded (using H264 or similar codecs), then encrypted (with RSA or similar standards), and finally stored across a distributed set of disks to ensure redundancy (e.g., RAID 5). These processes create a complex data-flow pipeline, differentiating two streams of video data: one for real-time inference and training, and another for archival. This dual-stream processing consumes considerable system resources such as CPU, memory, and energy, further complicating the management for intermittently powered systems like \US~\cite{usas} where data integrity and security must be maintained during power disruptions \footnote{Management and retrieval of data typically utilize a vector database like file-system, although this is beyond the scope of this discussion.}. Table~\ref{tab:ResUtil} provides an estimation of resource utilization on commercial systems underscoring the fact that compressing, encrypting and reliably storing the data, especially for (intermittent) edge servers is a bigger challenge in contrast to classical cloud storage servers.

\begin{table}[]
\resizebox{\columnwidth}{!}{%
\begin{tabular}{|c|l|l|l|ll|}
\hline
\multirow{2}{*}{Data} &
  \multicolumn{1}{c|}{\multirow{2}{*}{Task}} &
  \multicolumn{1}{c|}{\multirow{2}{*}{Algorithm}} &
  \% CPU   Utilization &
  \multicolumn{2}{c|}{\% DRAM Utilization} \\ \cline{4-6} 
 &
  \multicolumn{1}{c|}{} &
  \multicolumn{1}{c|}{} &
  \multicolumn{1}{c|}{16 Core Xeon} &
  \multicolumn{1}{c|}{Peak} &
  \multicolumn{1}{c|}{Average} \\ \hline
All                      & Encryptions & RSA512  & 2.18  & \multicolumn{1}{l|}{14.56} & 5.85  \\ \hline
All                      & Decryptions & RSA512  & 3.45  & \multicolumn{1}{l|}{17.2}  & 6.12  \\ \hline
\multirow{2}{*}{3D   PC} & Compression & OctTree & 26.78 & \multicolumn{1}{l|}{78.2}  & 32.54 \\ \cline{2-6} 
                         & Inflation   & OctTree & 29.24 & \multicolumn{1}{l|}{81.56} & 36.18 \\ \hline
\multirow{4}{*}{Video}   & Compression & ZStd    & 24.7  & \multicolumn{1}{l|}{62.54} & 24.5  \\ \cline{2-6} 
                         & Inflation   & ZStd    & 22.6  & \multicolumn{1}{l|}{79.18} & 29.43 \\ \cline{2-6} 
                         & Compression & H264    & 12.85 & \multicolumn{1}{l|}{52.46} & 21.4  \\ \cline{2-6} 
                         & Inflation   & H264    & 14.2  & \multicolumn{1}{l|}{69.46} & 26.18 \\ \hline
All                      & (un)RAID    & Unraid  & 11.25 & \multicolumn{1}{l|}{29.4}  & 19.24 \\ \hline
\end{tabular}%
}
\caption{Resource utilization while running different algorithms under classical data archival pipeline for multiple data modalities in an AWS h1.4xlarge storage-optimized instance.}
\label{tab:ResUtil}
\end{table}

\biul{The Challenges:}  In edge computing architectures, the \textbf{lack of data reuse} between the analytics and video data archival poses significant challenges. Classically, video data is streamed simultaneously to compute and storage systems for various processing tasks, such as inference, exemplar selection, and storage, thereby increasing I/O bandwidth and system processing demands. This processing complexity necessitates substantial compute and memory resources, escalating power consumption (e.g., systems with CPUs having a thermal design power of 145W and 64GB of DRAM as noted in Table~\ref{tab:ResUtil}) and requiring larger form factors, which exceed the capabilities of intermittent systems and challenge sustainability goals. Therefore, there is an urgent need to \textit{minimize compute and power requirements} in archival processes. Moreover, the divergence of analytics and archival pipelines from the outset suggests that optimizing both hardware, software and data-flow used in inference, learning and archival could significantly enhance throughput and energy efficiency. This can be achieved by using modern \textit{neural compression algorithms} instead of the classical encoding algorithm. However, the neural compression algorithm needs to be \textit{compute efficient} (reusing maximum analytics pipeline), have \textit{high compression ratio} (need to compete with H264) and \textit{feature rich} (could be decompressed and retrieved with a reasonable loss). Furthermore, it needs to take advantage of the \textit{data similarity between frames} to further minimize the storage footprint, and thereby reducing the form factor and need of frequent disk swapping/ maintenance. 

The second challenge comes from \textbf{privacy and security of sensitive data}. Considering the cheap commodity use storage devices are often plug and play, they are often vulnerable for data leak, especially if they are deployed in public, like urban mobility setting. Unlike secure data centers, these federated, distributed and public deployment could be susceptible to direct physical attacks for data breach. Although  modern encryption algorithms like RSA are secure, there is still a threat of \textit{store now decrypt later}\footnote{Given a powerful enough computer like quantum computers, RSA encrypted data can be decrypted, and therefore National Institute of Standards and Technology (NIST) called for proposals to develop post quantum cryptography algorithms~\cite{nistPQC}, and defined a standard for the same. One of the successful submissions -- which was later defined as a standard -- uses lattice-based encryption algorithm~\cite{lattice01} and will be the focus of our work.} kind of attack~\cite{SNDL1}. To mitigate this, \textit{quantum safe encryption algorithms needs to be used without hindering the throughput}. Furthermore, the design \textit{needs to be programmable} to ensure encryption keys to be changed regularly for additional security. 

\biul{Solution Space and Our Work:} This paper proposes \salt, a novel storage solution designed for  continuous learning edge servers by incorporating a hardware-software co-design framework that allows for efficient data archival and storage. \salt utilizes the state-of-the-art neural compression which partially uses the inference/ exemplar selection pipeline along with layered neural codecs to compress the video data. It also uses the motion vectors as a latent space to effectively use the inter-frame similarity, thereby further increasing the compression ratio. \salt also provides a hardware accelerated lattice-based quantum safe encryption mechanism. To tightly integrate these solutions to the storage space, while providing data security, \salt uses computational storage devices (CSDs)~\cite{xilinx} which reduce the energy consumption while keeping the compute pipeline unaltered. Our main \textbf{contributions} include:
\squishlist
\item We propose design of a hybrid storage pipeline equipped with computational storage drives (CSDs) where the different drives could communicate with each other in a peer-to-peer fashion. These CSDs synergistically orchestrate the archival related tasks between the storage controller CPU and the computational storage FPGAs. The hybrid storage system is capable of taking the computed frame features and the motion vectors from the compute hardware to perform a novel layered neural compression.

\item  We discuss the storage data-flow, the compute orchestration and mapping in the proposed system. This includes a hardware software co-design for compute-intensive applications along with mapping different functions to different hardware in the data pipeline. Furthermore, we add failure management support for the intermittent edge servers.

\item  We go beyond the compute and look into future-proofing the storage server by equipping it with quantum safe lattice-based encryption technique. We maximize the hardware utilization by reusing compute kernels from the compression pipeline. We detail the design of the hardware accelerated encryption and maximize the resource reuse between the exemplar selection and encryption.

\item Finally we perform an in-depth exploration of this integration, supported by real-world data from domains such as autonomous driving and urban mobility, to illustrate its effectiveness in continuous learning scenarios. The proposed design provides $\approx2.2\times$ latency and $\approx5.6\times$ data movement benefits compared to the state-of-the-art, on a single storage and $\approx4.8\times$ latency benefit in a multi-node system.
\squishend

% %-------------------------------------------------------------------------------

% %-------------------------------------------------------------------------------
%\vspace{-4pt}
\section{Background and Motivation}
\label{sec:bg-rw}
%%%%%
\subsection{Storage for Continuous Learning Edge Servers}
Recent developments in continuous learning for video analytics~\cite{ekya, usas, dacapo} has significantly boosted the capabilities and accuracy of learning systems. The major focus of these works have been building compute platforms with efficient scheduling~\cite{ekya, usas}, and reconfigurable hardware design~\cite{dacapo, usas}. This solves majority of the bottlenecks in a performance-driven classical cloud server platform. However, video analytics for many applications~\cite{Intelurbanmobility, googleUrbanMobility, AD01, AD02} are moving towards the edge. Therefore, managing and storing the hefty volume of video data brings more challenge due to the energy, compute and form-factor limitations. Modern and upcoming applications like urban mobility and autonomous driving are predicted to be generating hundreds of exabytes of data~\cite{UrbanTraffic, Intelurbanmobility, AD01, AD02} per year while increasingly being deployed at the edge calling for a robust, secure, and efficient infrastructure for storing data at the edge while needing occasional human intervention for maintenance. 

\biul{Using State-of-the-Art Video Data Storage? Maybe Not:} While storing video data as files works for small systems, in very large-scale systems they are typically stored using vector databases~\cite{vectorDB01, vectorDB02}. Vector databases typically extract features from the video data to form index and those indices are then sorted using various metrics like neighborhood, maximum similarity, etc.~\cite{indexingKD, indexingkmeans, indexingLHS, faiss} for faster retrieval. The vector index are used to point to the meta-data of the video file and then the actual video data is retrieved from the storage~\cite{vectorDB01}. This approach helps context-based search like looking for particular objects, events, or attributes~\cite{faiss}. However, it is obvious that this approach, albeit good for streaming and retrieval, are not entirely space-efficient, and at times can increase the data volume by many folds~\cite{faiss}. In an edge server where we are only worried about storage and not retrieval\footnote{We assume the data to be eventually available in the data repository, where they can be properly stored for efficient lookup. This can be done by periodically transporting the data by swapping out storage bays. Our goal is to maximize storage at edge so that the frequency of maintenance decreases.}, the vector database approach is {\em not} effective. Rather, storing the data in a compressed and encrypted format (with redundancies) is more efficient and therefore is the focus of our work.

\biul{Why Not the Usual Process?} Now that we know  Classical approach of video data storage involves  encoding and encrypting the video data before storing them in a redundant storage array~\cite{archivePipe01, archivePipe02, archivePipe03, archivePipe04} which consumes significant resources (refer Table~\ref{tab:ResUtil}). Even though there have been significant research in accelerating both compression~\cite{zstd, zstdfpga} and encryption~\cite{rsa, fastRSA01, fastRSA02}, operating on large-scale video data often demands more resource than what edge servers could afford~\cite{usas}. Co-locating this compute along with the inference and training would definitely hinder the critical path.

The main reason these algorithms consume significant resources is because of the amount of data they handle. Every single $1920 \times 1080$ raw frame prior to encoding carries $\approx 23$MiB of data which, @60fps, will require processing $\approx 1.4$GiB of data per second per imaging source. This pipeline assumes the use of classical data encoding algorithms like H264\footnote{We do {\em not} consider H265 here as currently in commercial systems H264 is the standard and typically enjoys hardware support. Moreover, H265 is an extension of H264 with additional features like coding tree units and intra-prediction directions which demand  significantly more computation.} which requires all the frames to encode a video stream, albeit it only saves the essential information. Therefore, it does not use any of the computations that are used in the inference and learning pipeline. Then, the question is: \textbf{is there a way we can reuse the computations used for the inference to help us in encoding the data?} The answer is  \textit{neural codecs}~\cite{neuralcodec01, neuralcodec02}.

\biul{Neural Codecs -- DNNs for Compression:} Neural codecs represent a paradigm shift in video compression technology, leveraging the capabilities of deep learning to optimize both encoding and decoding processes. Unlike traditional codecs that rely on predefined algorithms to compress video data, neural codecs utilize an end-to-end trainable system based on neural networks. These networks are trained on extensive video datasets, allowing them to dynamically adapt compression strategies based on the content's complexity and prevailing network conditions. The architectural backbone of neural codecs typically comprises an autoencoder, where the encoder compresses the video into a compact, lower-dimensional representation, and the decoder reconstructs it back into video format. These blocks can be stacked over each other to form layered codecs (like SHVC and SVC). This process benefits significantly from residual learning techniques, where each successive layer in the network aims to correct errors from the previous layers, thereby enhancing the reconstructed video quality incrementally with each additional decoding layer. 

Moreover, neural codecs excel in adaptability, offering robust performance across variable bandwidth and computational conditions, making them particularly suited for real-time streaming environments. These codecs can operate on generic computational hardware, such as GPUs and FPGAs, without the need for specialized video processing units, thus broadening their applicability across different device platforms. This adaptability also extends to content delivery dynamics, where neural codecs can adjust the streaming quality in real-time, responding adeptly to fluctuations in network throughput and variations in device capabilities. Such capabilities not only enhance user experience by minimizing buffering and maximizing video quality but also optimize bandwidth usage, presenting a cost-effective solution for content providers. These technical advancements position neural codecs as potential game-changers in the video streaming industry~\cite{DLSS}, promising significant improvements in efficiency, scalability, and flexibility in various streaming scenarios.

One major issue with neural codecs are their lack of utilization of inter-frame similarity. Classically, neural codecs compress each frame by treating them like images. However, video data often comes with a large amount of inter-frame similarity~\cite{frameSim, race2sleep, dejaview, holoar, pointcloud}. This similarity could be exploited, like in classical encoding algorithms, to further increase the compression ratio while improving the feature quality. Furthermore, neural codecs offer us the flexibility to approximate the computation (using quantization) to further enhance efficiency. Since they also use the computation blocks of the standard neural network, they can be jointly trained along with the classifier network to perform as the feature extraction and encoding backbone, thereby maximizing the utilization of the inference pipeline, and reducing the computation needed for compression. This \textbf{maximizes data and resource reuse}, and the pipeline for inference and compression do not diverge from the get going.

However, even with data reuse, performing compression using neural codecs would require extra compute resources, energy and latency, hindering the critical path, i.e., inference. However, \textit{this issue could be alleviated by not using the compute resource in the critical path and preferably moving the compute to the storage where the data will eventually be stored}. \textbf{This is where the modern and upcoming computational storage devices (CSDs) come to rescue: bringing computation closer to data while providing with maximum energy efficiency and programmability~\cite{samsungssd, xilinx}}. However, storage systems are not typically built to cater towards the ML applications, and now that compression becomes a ML application with the use of stacked neural codecs, building the right storage stack along with computational storage devices becomes an important problem.

\biul{Evolution of Computational Storage:} The advent of CSDs represents a paradigm shift, bringing computation closer to storage. These devices, by integrating CPUs or FPGAs into the storage medium, facilitate computation at the storage level. Initial applications of CSDs were confined to tasks like encryption/decryption, RAID, and compression. However, there has been a significant push towards enabling more complex workloads, including query processing on CSDs. Efforts to adapt CSDs for machine learning applications, albeit limited in scope to classical learning and data management, mark a crucial step forward. Yet, the expansion of these technologies to encompass large-scale storage stacks and the integration of CSDs into conventional storage systems, particularly for ML applications, remains a significant challenge and an open area of research.

\biul{Bridging the Gap in Storage System Design:} There is a discernible gap in the design and conceptualization of storage drives, systems and servers, especially in the context of ML applications. Historically, researchers have investigated these components individually, often focusing on high-performance computing, scientific computing, and database applications. However, the specific demands of ML applications on storage systems have largely been overlooked. storage stack architects often abstract the computational processes, neglecting considerations such as data movement cost, compute cost, and power requirements. Conversely, architects of storage drives, including CSDs, tend to overlook the broader application requirements, such as data prioritization, potential offloading of computational tasks to storage, and application-specific data compression and encryption strategies. Our research aims to {\em bridge} this gap, focusing specifically on the exigencies of continuous learning applications.

\subsection{The Problem: Understanding the Data Flow}
\noindent\textbf{Challenges in Data Movement:} In both consumer applications and high-performance computing (HPC) programs, the process of data collection followed by analytics is a critical operation~\cite{polardb, deepstore, csdBigData, CSDML1}. The efficiency of data movement significantly influences the performance and energy consumption of large-scale systems~\cite{kvCSD}. As earlier discussions highlighted, applications on edge servers need to store generated data on storage stacks. These data are can then be moved to a central facility where they can be further treated for efficient retrieval ans on demand streaming. 

\noindent\textbf{Urban Mobility -- A Case Study in Continuous Learning:} To thoroughly understand this data flow, let us examine ``urban mobility'', a prevalent continuous learning task~\cite{ekya}. This task dynamically adapts to changing traffic patterns and environmental conditions. In urban mobility, high-volume data streams from multiple sources, such as high-resolution traffic cameras, are first compressed then analyzed for exemplar selection. This process involves ``representation learning''~\cite{icarl}, where data is transformed into ``feature vectors'' using deep neural networks, followed by unsupervised learning techniques like k-means clustering. The goal is to identify unique or new classes of data for training while archiving known classes. In exemplar selection, the entire dataset is analyzed to detect classes with unique features, i.e.,  the images that are much different from the training data distribution or new classes that were not included in the training data. This is typically achieved through representation learning~\cite{icarl} where the data is first converted into a feature vectors using the convolution layers of a large DNN model (or multiples of them), and then performing an unsupervised learning based classification, e.g., k-means++~\cite{kmeanspp,kmeansppS}, to cluster the data. It is noteworthy that, the process of neural compression as well as inference/representation learning use the feature extraction method. In this work, our goal is to use this to our advantage and maximize the compute and data reuse.  

\textbf{\textit{The critical insight to be gleaned here is reusing data and compute pipeline prior to its transfer to storage could markedly diminish the costs associated with data movement. This approach would notably minimize the consumption of bandwidth and latency and curtail the compute requirements, thereby ensuring that the compute server's resources are judiciously employed only for indispensable computations. This strategic compute-reuse could help in optimizing the efficiency and efficacy of computational operations, especially in large scale data intensive and data driven applications.}}

\subsection{Why Not More Compute at Storage Stacks?}
Integrating additional compute capabilities within storage stacks to offload certain computational tasks may seem an intuitive solution to the problem at hand. Theoretically, storage stacks could handle operations like unRAID and decryption, and extend their functionality to data inflation and feature extraction. This would ostensibly allow for the selective transfer of only pertinent exemplar data to compute servers, thereby optimizing data movement costs. However, this approach presents several challenges. Current storage stacks are outfitted with robust CPUs and memory systems, which are heavily tasked with state-of-the-art storage management algorithms, as evidenced by the resource utilization outlined in TABLE~\ref{tab:ResUtil}. These algorithms already consume substantial compute and memory resources, often to the extent of fully occupying the storage controller system, with a portion of resources being allocated for essential system stack operations. Furthermore, incorporating more advanced compute hardware, such as FPGAs, GPGPUs and other accelerators,  would lead to underutilized I/O slots and memory, consequently escalating the cost of storage systems. Additionally, storage solutions that are optimized for I/O throughput tend to lose their specialized efficiency when repurposed for general compute tasks~\cite{vss}. While the previous research has suggested the idea of integrating analytics into storage systems, these discussions usually revolve around system architecture, scheduling, and data management, without delving into the requisite compute capabilities~\cite{vss, tasm, SSanalytics1, vstore}.

\subsection{More Compute in Storage? Why Not?}
Exploring the realm of computational storage drives presents a tantalizing avenue for on-disk computing capabilities. Storage controllers, typically constrained by I/O bandwidth, are now being complemented by the vast internal bandwidth of solid-state drives (SSDs), making them prime candidates for near-data processing. Recent advances in both commercial~\cite{samsungssd, xilinx, intelcsd, scalflux2, scalflux3, editicom, skssd} and academic sectors~\cite{csdhpc, wherecsd, csdppf, acchpc, nascent, nascent2, querycsd} advocate the use of computational storage drives (CSDs) across databases, high-performance computing (HPC), and analytics. AMD and Xilinx have introduced specialized tools and libraries designed to harness CSDs~\cite{vitis, xrt}, enabling peer-to-peer PCIe transactions that bypass the CPU~\cite{xrt}. The emergence of Compute Express Link (CXL) technology~\cite{cxl, cxl1, cxlssd} further amplifies the potential of disaggregated storage and memory systems. 

Decoupling compute tasks needed for storing the data from the host CPU and embedding them directly within storage devices, particularly through CSDs, has demonstrated significant performance and energy benefits. Tasks traditionally performed at the storage controller level are now being offloaded to CSDs, often accelerated using FPGA primitives~\cite{understandingCSDs, xrt, vitis}. Moreover, CSDs hold the potential to undertake critical machine learning tasks like feature extraction and clustering, streamlining tasks like neural compression. The unique combination of FPGAs' parallel processing prowess, the substantial internal bandwidth of SSDs, and their block-accessible nature position CSDs as ``ideal components'' for evolving smart storage solutions tailored for machine learning. 

To cater towards this, in this work, we propose \salt --  a mini computational storage server (we call it ``edge storage server'') stack for managing the data archival in edge servers. \salt provides adaptive data compression using neural codecs and further enhances the data security by providing an accelerated quantum safe data encryption policy, protecting these vulnerable edge storage servers against the store now decrypt later~\cite{SNDL1} attacks.

%%%%%

% %-------------------------------------------------------------------------------

% %-------------------------------------------------------------------------------
%\vspace{-4pt}
\section{Data Compression using Neural Codec}
\label{sec:dse}
%%%%%
In this section we go over the overall design and design choices for the neural codec design of \salt. We discuss the edge storage architecture followed by the choice of neural codec and their design.

\subsection{\salt Storage Architecture:} 
\begin{figure}[!htbp]
\centering 
\includegraphics[width=0.7\linewidth]{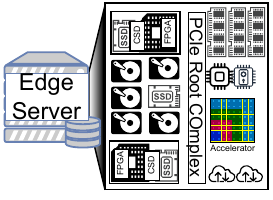}
\caption{High-level design of the \salt edge server - it consists of the accelerated video analytics compute along with computational storage and classical storage drives.}
\label{fig:saltLg}
\end{figure}

\salt edge storage architecture is crafted to optimize data-flow and computational efficiency in continuous learning edge servers. \salt employs a ``hybrid model'', integrating Computational Storage Drives (CSDs) with Field-Programmable Gate Arrays (FPGAs)~\cite{xilinx} and classical storage drives. This design, depicted in Fig.~\ref{fig:saltLg}, gives the programmable computational capabilities of CSDs along with the cost-effectiveness and durability of the classical HDDs. It leverages the PCIe interface for efficient ``peer-to-peer communication'', substantially reducing communication latency and energy requirements. This setup alleviates the host CPU from handling frequent data movement interruptions. A storage platform entirely designed with CSDs is not pragmatic at the current time because of their exorbitant cost and power consumption~\cite{xilinx, polardb} . However, a combination of CSDs with classical storage medium provides the most optimal solution and hence motivates our design. 

\biul{Data-flow Reorganization in \salt}: At the core of \salt's design is the ``data-aware'' reorganization of compute processes. Unlike conventional storage servers, \salt discerns between the ``data path''  and the ``resource path'' for system I/O calls, translating these requests into CSD-specific functions. These functions leverage FPGA kernels for efficient data processing.

As discussed earlier,\salt edge storage implements a video archival by using neural compression followed by a quantum safe encryption (refer Fig.~\ref{fig:saltpipe}). To maximize the compute reuse between the compute pipeline and the archival pipeline, \salt uses apart of the neural network of the inference engine to extract features, and then further performs the encoding using the FPGA in the CSD. 

\begin{algorithm}
\caption{Neural Encoding and Compression using the video data inference pipeline.}
\begin{algorithmic}[1]
\State \textbf{Input:} Video frames sequence $F = \{f_1, f_2, \dots, f_n\}$
\State \textbf{Output:} Compressed frames $C = \{c_1, c_2, \dots, c_n\}$

\State Initialize MobileNet Model $M$
\State Initialize Autoencoder Decoder $D$
\State Initialize Motion Vector Extractor $V$

\Procedure{ExtractFeatures}{$frame$}
    \State $features \gets M(frame)$ \Comment{Extract features using MobileNet}
    \State \textbf{return} $features$
\EndProcedure

\Procedure{CompressFeatures}{$features$}
    \State $compressed \gets D(features)$ \Comment{Compress using autoencoder}
    \State \textbf{return} $compressed$
\EndProcedure

\Procedure{CalculateMotionVectors}{$frame_{current}, frame_{previous}$}
    \State $motion\_vectors \gets V(frame_{current}, frame_{previous})$
    \State \textbf{return} $motion\_vectors$
\EndProcedure

\Procedure{StackCompression}{$current\_compressed, motion\_vectors$}
    \State \textbf{return} some compression algorithm using $current\_compressed$ and $motion\_vectors$
\EndProcedure

\For{$i \gets 1$ \textbf{to} $n$}
    \State $features_i \gets \Call{ExtractFeatures}{f_i}$
    \State $c_i \gets \Call{CompressFeatures}{features_i}$
    
    \If{$i > 1$}
        \State $m_i \gets \Call{CalculateMotionVectors}{f_i, f_{i-1}}$
        \State $c_i \gets \Call{StackCompression}{c_i, m_i}$
    \EndIf
    \State $C[i] \gets c_i$
\EndFor

\end{algorithmic}
\label{algo:codec}
\end{algorithm}

%%%%%%%%%%%%%%%%%%%

%%%%%%%%%%%%%%%%%%%
\begin{algorithm}
\caption{Training Auto-Encoder with Motion Vectors and Stacked Compression while freezing the inference model.}
\begin{algorithmic}[1]
\State \textbf{Input:} Set of training video sequences $V$
\State \textbf{Initialize:} MobileNet $M$ \Comment{Weights frozen}
\State \textbf{Initialize:} Autoencoder $A$ \Comment{Trainable}
\State \textbf{Initialize:} Motion Vector Extractor $V$

\Procedure{ExtractMotionVectors}{$frame_{current}, frame_{previous}$}
    \State $motion\_vectors \gets V(frame_{current}, frame_{previous})$
    \State \textbf{return} $motion\_vectors$
\EndProcedure

\Procedure{ForwardPass}{$video$}
    \State $previous\_features \gets \text{null}$
    \State $previous\_compressed \gets \text{null}$
    \For{each frame $frame$ in $video$}
        \State $features \gets M(frame)$ \Comment{Extract features using frozen MobileNet}
        \State $compressed \gets A.encode(features)$ \Comment{Compress features}

        \If{$previous\_compressed \neq \text{null}$}
            \State $motion\_vectors \gets \Call{ExtractMotionVectors}{frame, previous\_frame}$
            \State $stacked\_input \gets \text{concatenate} \newline
            (compressed, previous\_compressed, motion\_vectors)$
            \State $compressed \gets A.reencode(stacked\_input)$ \Comment{Stacked compression}
        \EndIf

        \State $reconstructed \gets A.decode(compressed)$ \Comment{Decompress to reconstruct}
        \State Calculate reconstruction loss between $frame$ and $reconstructed$
        \State $previous\_frame \gets frame$
        \State $previous\_compressed \gets compressed$
        \State $previous\_features \gets features$
    \EndFor
    \State Backpropagate loss and update weights of $A$ only
\EndProcedure

\While{not converged}
    \For{each $video$ in $V$}
        \Call{ForwardPass}{$video$}
    \EndFor
\EndWhile

\end{algorithmic}
\end{algorithm}
%%%%%%%%%%%%%%%%%%%
Neural codecs present a new paradigm video compression technology, leveraging deep learning to surpass traditional codec efficiency~\cite{neuralcodec01, neuralcodec02}. Traditional neural codecs, while innovative, typically encode and decode video streams in a monolithic fashion, which often results in suboptimal utilization of computational resources and inflexibility~\cite{neuralcodec01}. This inherent inefficiency stems from their design which does not allow incremental improvements in video quality and often leads to either over-utilization or under-utilization of bandwidth. To address these shortcomings, the advent of layered neural codecs marks a significant advancement~\cite{swift}. Layered neural codecs, by design, encode video into multiple, distinct layers of data, each enhancing the video quality incrementally. This allows for dynamic adaptation to network fluctuations and client-side computational capabilities, thereby optimizing the streaming experience. The capability to adjust the quality dynamically, by decoding additional layers as resources permit, ensures efficient bandwidth usage and enhances the overall Quality of Experience (QoE), making layered neural codecs a compelling choice for modern video streaming solutions.

In an effort to optimize video compression beyond traditional and current neural codec capabilities, we introduce an enhanced layered neural codec that utilizes both intra-frame and inter-frame redundancies. Our approach extends the layered neural codecs by incorporating motion vectors as a latent space, akin to the macroblock techniques used in H.264, to maximize inter-frame compression efficiency. By exploiting the temporal correlations between consecutive frames, our codec can significantly reduce the required bitrate while maintaining high video quality.

The primary innovation lies in leveraging motion vectors to identify and encode only the changes between frames, rather than re-encoding entire frames. This technique, combined with the use of anchor frames—similar to keyframes in traditional codecs—allows the codec to understand and predict frame sequences more effectively, reducing redundancy and enhancing compression.
Let $F_t$ represent the frame at time $t$, and $F_{t-1}$ be the anchor frame. The motion vector field $M_t$ between $F_t$ and $F_{t-1}$ is computed to capture the displacement of pixels. Each frame $F_t$ is divided into blocks $B_{t,i}$, where $i$ indexes the block within the frame.
The residual frame $R_t$ for each frame is calculated as:
\[
R_t = F_t - \text{predict}(F_{t-1}, M_t)
\]
where $\text{predict}(\cdot)$ is a function that reconstructs $F_t$ from $F_{t-1}$ using the motion vector field $M_t$. This prediction involves translating the blocks of $F_{t-1}$ according to $M_t$ and serves as the predicted frame. The residual frame $R_t$, which contains only the differences not captured by the motion prediction, is then encoded using the layered neural network.
Each layer of the neural network encodes progressively finer details of $R_t$. If $L_k(R_t)$ represents the $k$-th layer's encoding of the residual frame, the overall encoding of the frame can be expressed as:
\[
E_t = \sum_{k=1}^K L_k(R_t)
\]
where $K$ is the total number of layers, and each $L_k$ encodes different levels of detail or different regions of the frame, based on the motion information and the prediction error. Algorithm~\ref{algo:codec} shows the details of the layered neural codec. 

We implement this neural codec using the FPGA in the CSDs. FPGAs are ideal for this application due to their parallel processing capabilities and the ability to handle multiple data streams concurrently. The implementation of layered codecs involve the following components.\textbf{1. Motion Estimation:} Utilize dedicated hardware blocks for calculating motion vectors between consecutive frames. This step can leverage FPGA's DSP slices for fast cross-correlation or block matching algorithms. \textbf{2. Prediction and Residual Calculation:} Implement pipelined architectures for the $\text{predict}(\cdot)$ function and subsequent residual calculation to minimize latency. \textbf{3. Layered Encoding:} Each layer of the neural codec can be implemented using parallel processing units in FPGA, allowing simultaneous processing of different frame parts or different quality layers. \textbf{4.Data Flow Management:} Design efficient data paths to handle the high throughput of video data and intermediate results between the FPGA blocks, ensuring that bandwidth and memory access bottlenecks are minimized.
By optimizing each component for FPGA execution and taking advantage of the hardware's ability to execute multiple operations in parallel, the proposed codec can achieve real-time performance even for high-resolution video streams. The use of FPGA not only accelerates the processing speed but also provides flexibility to adapt to various codec configurations.

\biul{Training the Neural Codecs to Utilize Inference Pipeline:} The enhancement of our layered neural codec involves a joint training regimen that integrates the model used in inferecne pipeline, specifically MobileNet, as a static feature extractor within the compression framework. This strategy capitalizes on the robust, pre-trained features of MobileNet, which are frozen during training to ensure their integrity and to leverage their proven capability in capturing essential visual features. The extraction of motion vectors between frames further enriches the feature space by incorporating temporal dynamics essential for effective compression. 

The codec's autoencoder component, which is trainable, is then tasked with compressing these enriched features. This setup not only streamlines the encoding process by utilizing high-quality features but also significantly enhances compression efficiency by exploiting both intra-frame richness and inter-frame continuity. The training process is designed to optimize the autoencoder's ability to compress and decompress video sequences efficiently, without altering the pretrained feature extractor, thereby providing a stable, high-performance baseline for feature representation. The mathematical formulation for this training process is centered around minimizing the reconstruction loss, $L = \sum_{t=1}^{N} \| F_t - \hat{F}_t \|_2^2$; where $F_t$ is the original frame at time $t$, and $\hat{F}_t$ is the reconstructed frame, obtained by decoding the compressed representation that was encoded using features extracted via MobileNet and refined by the motion vector-informed autoencoder. The backpropagation is applied only to the layers of the autoencoder, ensuring that the feature extractor's parameters remain intact. This approach not only preserves the integrity of the visual features derived from MobileNet but also tailors the compression mechanism to be highly adaptive to the content-specific characteristics captured by these features.

%%%%%

% %-------------------------------------------------------------------------------

% %-------------------------------------------------------------------------------
%\vspace{-4pt}
% \section{Proper Exemplar Selection in Machine Learning Storage Servers}
% \label{sec:Exm}
% \input{src/04ExemplarSelction}

% %-------------------------------------------------------------------------------

% %-------------------------------------------------------------------------------
%\vspace{-4pt}
\section{Optimizing Lattice-Based Cryptography}
\label{sec:SE}
Followed by the video compression, in this section, we discuss the quantum safe encryption technique. Among quantum safe encryption algorithms, Lattice-based cryptography (LBC)~\cite{lattice01, lattice02}, grounded in hard lattice problems, offers robust resistance against quantum attacks. This quantum resilience is vital for protecting large data-sets on machine learning storage servers, particularly from 'store now, decrypt later' threats.

\begin{algorithm}[!htbp]
\caption{Lattice-Based Encryption Process}
\label{alg:lattice_based_encryption}
\begin{algorithmic}[1]
\Procedure{Lattice\_Based\_Encrypt}{message, public\_key}
    \State PM $\gets$ ConvertToPolynomial(message)
    \State (PA, PE) $\gets$ GenerateRandomPolynomials()
    \State C1 $\gets$ PolynomialMultiply(PA, public\_key) \Comment{Utilizing HSPM}
    \State C2 $\gets$ PolynomialMultiply(PA, PM) \Comment{Employing SDMM}
    \State \textbf{return} (C1, C2)
\EndProcedure
\end{algorithmic}
\end{algorithm}

At the heart of LBC lies the Ring-Learning with Errors (R-LWE) algorithm, which translates plaintext messages into polynomial representations and intertwines them with random polynomials using complex multiplications and additions, as delineated in Algorithm~\ref{alg:lattice_based_encryption}. Note that, within the LBC algorithm, certain components, specifically polynomial multiplications, exhibit similarities to operations performed in CNNs. These similarities open up the possibility of reusing hardware designed for CNN operations to accelerate LBC computations. The most common algorithms used fod the said operations are High-Speed Schoolbook Polynomial Multiplication (HSPM) and Pipelined Systolic Dimension Modular Multiplier (SDMM), and we propose to accelrate the same using the FPGAs in the CSDs. The data locality 
 due to the SSD and the high throughput due to the FPGA could facilitate swift polynomial processing and refined modular multiplications using DSP slices, thereby accelerating the encryption process.

\begin{figure}[!htbp] 
 \centering
    \subfloat[HSPM micro-architecture.]{\includegraphics[width=0.7\linewidth]{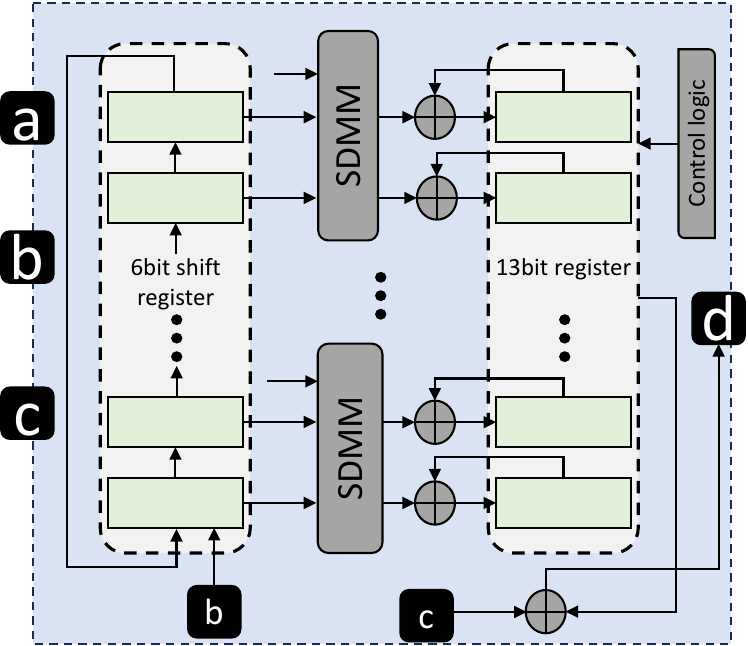}\label{fig:hspm}}

    \subfloat[Pipeline SDMM Structure micro-architecture.]{\includegraphics[width=0.7\linewidth]{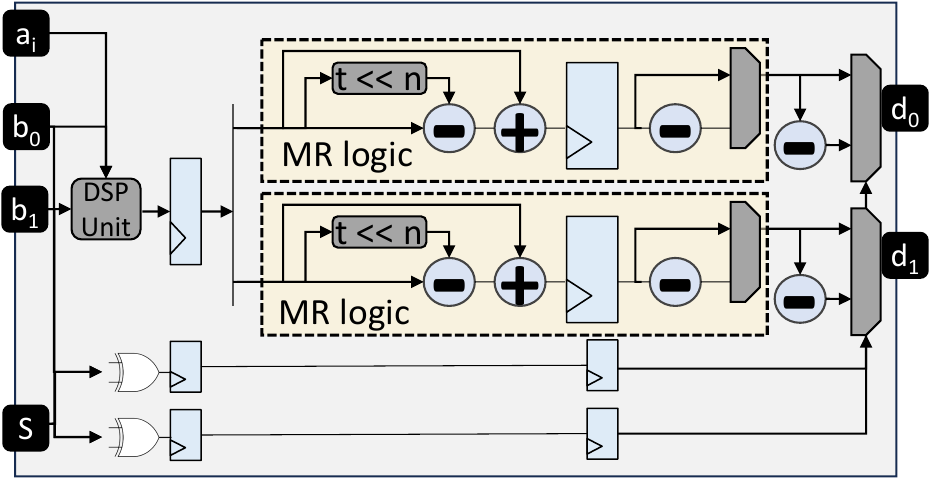}\label{fig:sdmm}}
    \caption{Microarchitectural designs of HSPM and SDMM: Hardware modules for polynomial multiplication in LBC.}
    \label{fig:hardware_microarchitectures} 
\end{figure}

\biul{Designing HSPM Accelerator on CSD FPGA: }
The HSPM hardware is characterized by its fully parallelized design, incorporating 128 Multiply-Accumulate (MAC) units for handling polynomials of degree \( n = 256 \). Each MAC unit within the HSPM architecture is capable of conducting two parallel modular multiplications. This is achieved through the use of a single Digital Signal Processing (DSP) block that operates on signed data representation. Consequently, these units are referred to as Signed Double Modular Multiplication (SDMM) units. 
The HSPM accelerator's architecture, as illustrated in Fig.~\ref{fig:hspm}, comprises three pipelined stages. These stages include the data loading phase, the modular polynomial multiplication via the SDMM unit, and the final accumulation registers. Data is input into the HSPM in a serial-to-parallel conversion process, while the outputs, after undergoing addition operations, are retrieved serially through an address signal.

Central to the Ring-Learning with Errors (R-LWE) based Public Key Encryption (PKE) is the equation \( d = a \cdot b + c \). In this context, the operand \( a \) can represent polynomials such as \( p \), \( a \), or \( c_1 \) from Algorithm 1, while \( b \) corresponds to \( e_1 \), \( r_2 \), and \( c \) to \( e_2 \), \( e_3 \), \( c_2 \). During the data loading phase, the 256 coefficients of polynomial \( b \) are input serially into a 6-bit shift register. Simultaneously, the first coefficient \( a_0 \) of polynomial \( a \) is fed in parallel to all 128 SDMM units. Post-loading, each SDMM unit commences the modular multiplication of each coefficient \( a_i \) with the entirety of \( b \)'s coefficients in parallel. This process repeats for all coefficients \( a_i \).
%, ranging from 0 to 255.
The resultant outputs from each SDMM calculation are accumulated every cycle. The final polynomial product \( p \) is sequentially read out by the address signal \( \text{addr}_{\text{ab}} \), combined with the coefficient of \( c \), thereby producing the final output \( d \) as \( d = a \cdot b + c \). This output is then transmitted serially over \( n \) clock cycles.

\biul{Designing SDMM Hardware on CSD FPGA: }
The SDMM hardware is innovatively designed to perform two modular multiplications per DSP Slice. This is achieved through the implementation of signed Gaussian sampling~\cite{liu2019optimized}, for the error-vector \( e_1 \) and the secret-key \( r_2 \). In this signed representation, the Gaussian distribution ranges from \( [0, ks) \) to \( [q - ks, 0) \), where \( k \) takes integer values \( 1, 2, 3, \ldots \), and is symmetrically distributed around \( m = 0 \). Consequently, there is no data in the range \( [0, q - 1) \), allowing the maximum value of the 13-bit samples to be represented efficiently by a 6-bit signed number. The modular multiplication utilizing these signed numbers is formulated as follows:
\begin{equation}
b \otimes a \mod q = q - [(b - a) \mod q]
\end{equation}
Here, the most significant bit (MSB) signed-bit dictates the subtraction of the result from the modulus \( q \), as illustrated in Fig.~\ref{fig:sdmm}. 
A single DSP unit includes a multiplier, with the lowest 18 bits of the DSP output representing the first product \( d_0 \) and the highest 18 bits indicating the second product \( d_1 \), corresponding to \( b_0 \) and \( b_1 \) respectively, as depicted in Fig.~\ref{fig:sdmm}.

Following the multiplication, the two 18-bit products undergo separate modular reductions. A Modular Reduction (MR) circuitry based on approximation~\cite{kundi2020axmm}. Our MR unit achieves consumes $\approx 82\%$ less hardware compared to classical implementation. This process is constant time, requiring only a single subtraction of \( q \). The MR hardware, highlighted in Fig.~\ref{fig:sdmm}, consists of a single shift block, a subtractor, and a adder, consuming much less hardware then the state-of-the-art~\cite{fan2018lightweight}. The MR, in conjunction with the multiplier, delivers the output within 2 clock cycles. The SDMM also offers a simpler controller design for multiplication and in-place product term reduction after vector-vector multiplications. Notably, the SDMM can be easily reconfigured to support any modulus.

%%%%

% %-------------------------------------------------------------------------------

% %-------------------------------------------------------------------------------
%\vspace{-4pt}
% \section{Integration and Design Implementation}
% \label{sec:imple}
% \input{src/05DesignImplementation}

% %-------------------------------------------------------------------------------

% %-------------------------------------------------------------------------------
%\vspace{-4pt}
\section{Implementation and Evaluation  }
\label{sec:eval}
%%%%%

To implement and evaluate \salt, we chose two different platforms -- 1) Using a server-grade system with two Xilinx-Samsung computational storage drive~\cite{xilinx}, and  2) Using amazon web services (AWS) F1 instances, which have AMD Alveo FPGA which can work as the compute capable of peer-to-peer communication and thereby enabling a computational storage platform. We use  Vitis 2022.2 along with Xilinx Runtime Library~\cite{xrt} for programming the FPGAs in both computational storage drives and the Alveo card. The configuration of the server with Xilinx CSD has a 12-core Xeon bronze CPU with 128GB memory, $2\times$3.84TB CSDs, and $2\times$2TB SSDs. Similarly, the AWS server has 24 cores, with 192GB memory, one Alveo FPGA and 2TB SSDs.

\begin{figure}[h]
\centering 
\includegraphics[width=0.7\linewidth]{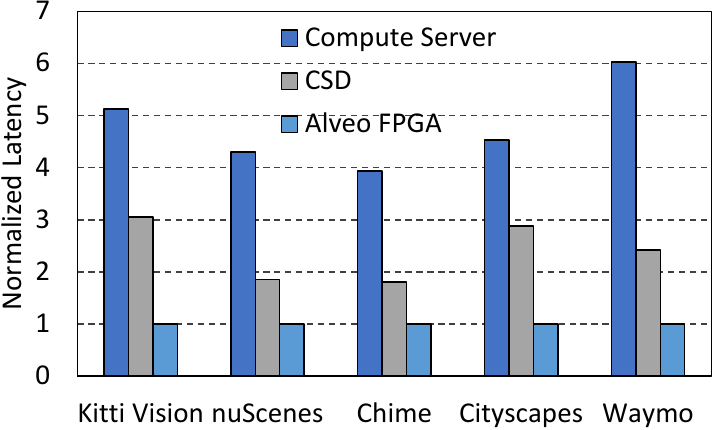}
\caption{Latency analysis of \salt on the commercial Xilinx CSD on a workstation class machine (lower is better). Compute server indicates a software only classical storage solution without CSDs.}
\label{fig:csd}
\end{figure}

\begin{table}[h]
\resizebox{\columnwidth}{!}{%
\begin{tabular}{|l|l|l|}
\hline
\multicolumn{1}{|c|}{\textbf{Data   Location}} & \multicolumn{1}{c|}{\textbf{kernel Execution}} & \multicolumn{1}{c|}{\textbf{SpeedUp}} \\ \hline
CSD1                      & CPU                       & 1     \\ \hline
CSD1                      & CSD1                      & 3.9   \\ \hline
CSD1   (0.1X), CSD2(0.9X) & CSD1   (0.1X), CSD2(0.9X) & 4.46  \\ \hline
CSD1   (0.3X), CSD2(0.7X) & CSD1   (0.3X), CSD2(0.7X) & 5.608 \\ \hline
CSD1   (0.4X), CSD2(0.6X) & CSD1   (0.4X), CSD2(0.6X) & 6.67  \\ \hline
CSD1   (0.5X), CSD2(0.5X) & CSD1   (0.5X), CSD2(0.5X) & 7.7   \\ \hline
\end{tabular}%
}
\caption{Effect of Data distribution on compute speed-up.}
\label{tab:CSDDataDistribution}
\end{table}

\begin{figure*}[!h] 
 \centering
    \subfloat[Accuracy.]{\includegraphics[width=0.31\linewidth]{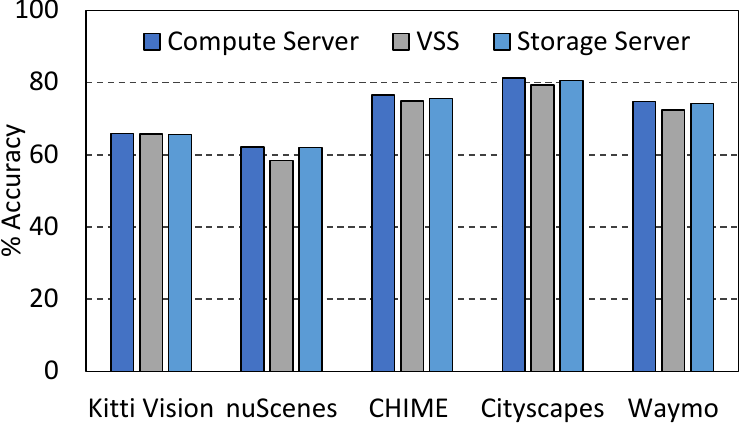}\label{fig:accuracy}} \hfill 
    \subfloat[Latency.]{\includegraphics[width=0.3\linewidth]{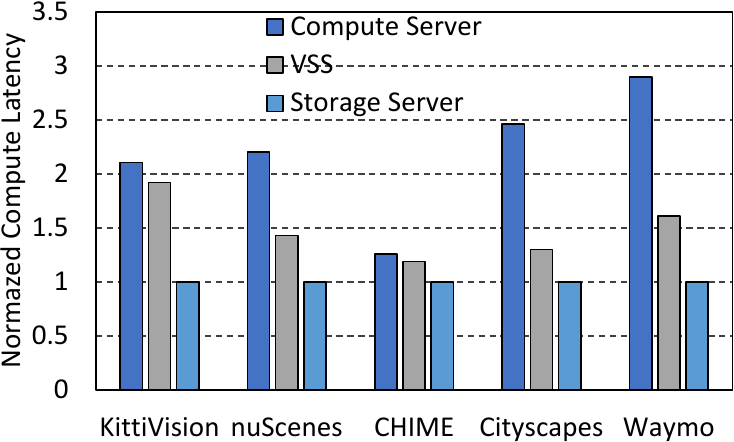}\label{fig:latency}} \hfill
    \subfloat[Data Volume.]{\includegraphics[width=0.3\linewidth]{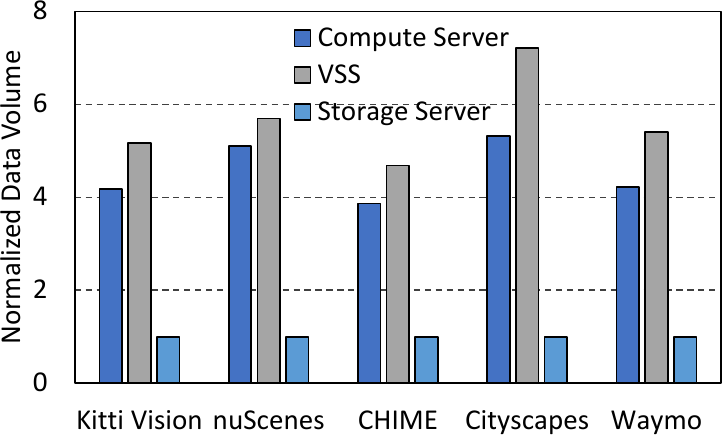}\label{fig:datavolume}}
    \caption{Performance of \salt on larger compute and storage nodes. This experiment to mimics a consolidated edge server catering to many video streams. Compute server indicates a software only classical storage solution without CSDs.}
    \label{fig:pubcloud} 
\end{figure*}

\subsection{Evaluation of Encoding, Scaling and Accuracy}
In evaluating the \salt storage system, particularly for continuous learning scenarios video analytics applications, we selected data-sets that are not only dense but also necessitate continuous learning due to their dynamic nature. Autonomous driving and urban mobility applications, generating over 400TB of data annually~\cite{AD01, AD02, ekya}, predominantly comprise video and 3D point cloud data, making them ideal for our evaluation. The \textit{Cityscapes} data-set~\cite{cityscapes} is a comprehensive collection of urban street scenes from 50 different cities, providing a rich source of annotated video data for semantic urban scene understanding. This data-set is particularly suited for evaluating the performance of \salt in dense, urban environments. Another video data-set, the \textit{Waymo Open Data-set}~\cite{waymo}, is one of the largest and most diverse data-sets for autonomous driving. It offers high-resolution sensor data, including LIDAR and camera recordings, across a variety of urban and suburban landscapes. This data-set's volume and diversity make it an excellent benchmark for assessing \salt's capabilities in handling large-scale, real-world data. For 3D point cloud data, we selected the \textit{KITTI Vision Benchmark Suite}~\cite{kitti}, a fundamental data-set in autonomous driving research. It includes a variety of data types and tasks, providing a real-world urban driving context that is essential for testing \salt's performance in processing and managing 3D spatial data. Additionally, the \textit{nuScenes} data-set~\cite{nuscenes} by Aptiv, with its comprehensive sensor data and annotations covering diverse driving scenes, is instrumental in evaluating \salt's efficiency in multi-modal data processing typical in urban mobility scenarios. Beyond the vision-based data, we also incorporated the \textit{Chime Audio} data-set~\cite{chime} into our evaluation. This data-set, consisting of audio recordings, offers a different modality to test the versatility of \salt in handling various types of continuous learning data beyond visual inputs. 
In our comparative analysis, we employ \textit{VSS: A Video Storage System}~\cite{vss} as a ``baseline'', given its innovative approach in optimizing video data management. VSS excels in decoupling high-level video operations from storage and retrieval processes, efficiently organizing data on disk, enhancing caching mechanisms, and reducing redundancies in multi-camera setups~\cite{vss}. This system has demonstrated significant improvements in VDBMS read performance, up to 54\%, and a reduction in storage costs by up to 45\%~\cite{vss}. VSS's emphasis on video data optimization makes it a pertinent benchmark for assessing the \salt storage system's capabilities in managing large-scale machine learning data-sets. 

We first implemented \salt on a workstation-class machine with two Xilinx CSDs - which colosely mimics the classical edge server setup. Fig.~\ref{fig:csd} shows the latency while using the storage server and CSD for performing data writes, {\em normalized} to the latency of doing the same using the state of the art Alveo FPGAs. \salt on CSDs perform $\approx1.99\times$ better than the implementation on classical storage systems. To further mimic realwold scenarios of multiple cameras sending multiple streams with various stream rate, we {\em distributed} video data into across two CSDs in different ratio, as shown in TABLE~\ref{tab:CSDDataDistribution}. Trivially, a load-balanced scenario outperforms any of the biased data distribution scenario. However, it is evident that any compute offloaded to any of the CSDs in these scenarios offers benefits in terms of data processing latency.

\begin{figure}[ht]
\centering 
\includegraphics[width=0.7\linewidth]{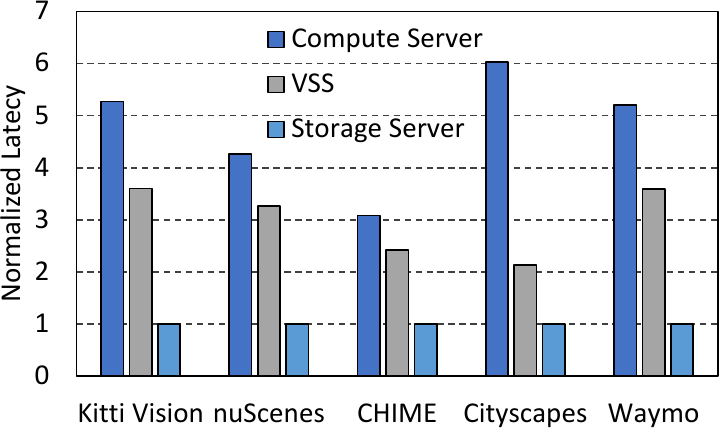}
\caption{Impact of scaling to multiple storage nodes.}
\label{fig:multinode}
\end{figure}

To underscore the impact of \salt at a much larger scale, where one can assume a consolidated edge server catering towards multiple streams as depicted in Ekya~\cite{ekya}, we utilized an AWS EC2 F1 instance with Alveo FPGAs, along with EC2 P4 instance with A100 GPUs to implement the continuous learning end-to-end. As shown in Fig.~\ref{fig:accuracy} \salt keeps up with the accuracy of the traditional compression system thanks to the robust layered coding algorithm. Additionally, as shown in Fig.~\ref{fig:latency}, \salt has $4.49\times$ less latency than VSS where as it shows about a $6.18\times$ speedup compared to the classical storage server. Furthermore, as depicted in Fig.~\ref{fig:datavolume}, SLAT reduces the data communication volume up to $\approx5.63\times$ compared to the classical approaches, thereby saving bandwidth (reduces bandwidth by $\approx36\%$, public cloud has strict regulations on measuring the actual bandwidth, and hence we report the approximate result from the traffic pattern) and energy. 

\begin{figure}[!ht]
\centering 
\includegraphics[width=0.7\linewidth]{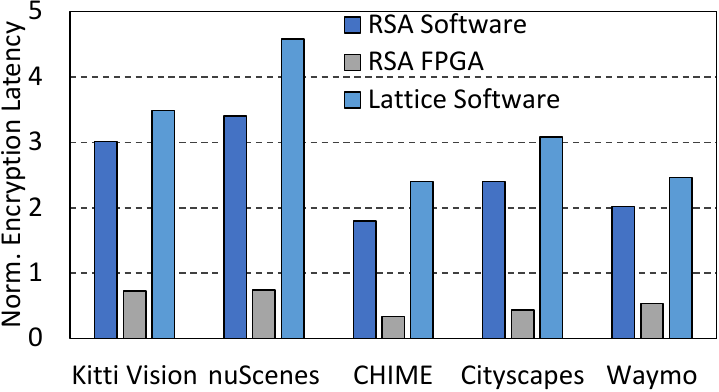}
\caption{Proposed encryption vs the state-of-the-art normalized to the FPGA implementation of the lattice based encryption.}
\label{fig:encryptioneval}
\vspace{-4mm}
\end{figure}

Since majority of the storage systems are not limited to one storage server, but are spread across multiple servers, we scaled \salt by deploying it in a distributed fashion. We used $5\times$ AWS EC2 F1 instances with Alveo FPGAs as storage nodes, along with one EC2 P4 instance with A100 GPUs as the compute server. We do not observe much change in the data volume. However, due to complex interaction between multiple storage nodes, especially where some of the data needed to arrive at different nodes via network, in Fig.~\ref{fig:multinode}, we observe a significant change in the latency compared to a single storage node. Although, a multi-node setup provides more parallelism, the speedup is sub-linear, and we observe $\approx3\times$ and $\approx4.77\times$ speedup against VSS and a classical storage server, respectively. 

\subsection{Evaluation of Video Data Recovery and Quality:}
\begin{figure}[!ht]
\centering 
\includegraphics[width=0.7\linewidth]{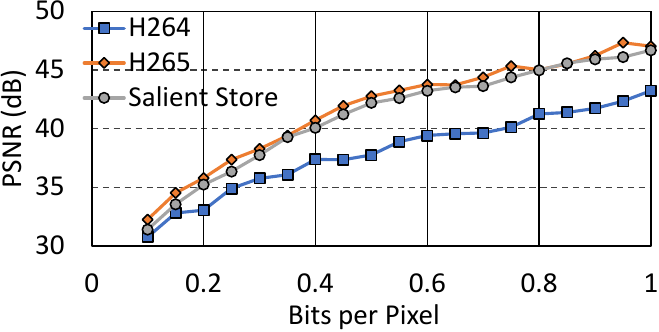}
\caption{Compression and Recovery efficiency.}
\label{fig:psnr}
%\vspace{-8mm}
\end{figure}

\begin{figure}[!ht]
\centering 
\includegraphics[width=0.7\linewidth]{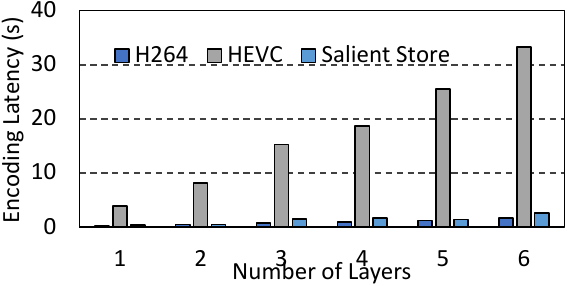}
\caption{Encoding Latency using layered coding.}
\label{fig:abalation}
%\vspace{-8mm}
\end{figure}

To ensure proper video quality upon recovery, we perform a peak signal-to-noise ratio (PSNR) study of \salt with the classical encoding mechanisms H264~\cite{h264} and H265 (HEVC)~\cite{h265}. Experiments on waymo~\cite{waymo} dataset shows the PSNR of \salt compared to the classical H264 and HEVC encoding pipeline in Fig.~\ref{fig:psnr}. While \salt is consistently performing better than H264, HEVC thanks to it's novel approach of fine grained computation at times outperforms our approach. A similar trend is observed in other datasets. With complex and high dimensional video data, HEVC computation complexity increases exponentially, and is more pronounced because of lack of hardware support. However, it is noteworthy that at higher bitrates, \salt consistently achieves high quality recovery with PSNR reaching $\approx47$dB. Furthermore, as we can see in Fig.~\ref{fig:abalation}, HEVC takes significantly more latency compared to \salt and therefore not entirely suitable for high frame-rate applications with resource constraints.

\subsection{Evaluation of Lattice Based Encryption}
One of the major contribution of \salt is accelerating the lattice-based encryption with the help of FPGAs on the storage nodes. Fig.~\ref{fig:encryptioneval} shows the comparison of our FPGA-accelerated lattice-based encryption against other state-of-the-art-techniques. RSA, the most popular encryption algorithm, when implemented using FPGA, outperforms our proposed hardware solution. However, our proposed solution offers $\approx3.2\times$ speedup compared to its software counterpart and a $\approx2.5\times$ speedup compared to the software based RSA algorithm. While we  do agree that the lattice-based encryption has more overheads compared to the RSA algorithm, the benefit of being ``quantum-safe'' outweighs the minimal cost incurred on encryption.  

\begin{figure}[h]
\centering 
\includegraphics[width=0.7\linewidth]{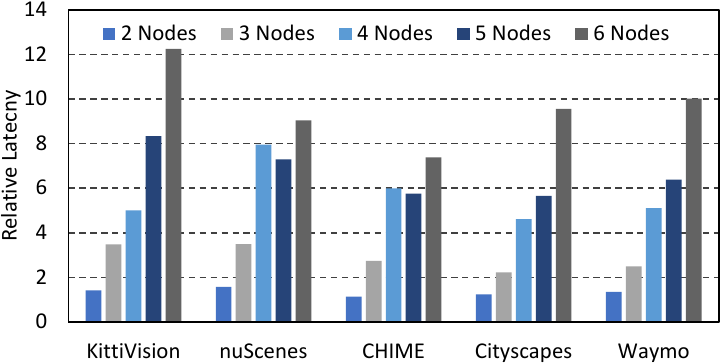}
%\vspace{-6pt}
\caption{Change of data movement latency with respect to the number of storage servers.}
\label{fig:nodeVlatency}
\end{figure}

Fig.~\ref{fig:nodeVlatency} shows the impact of scaling \salt with respect to a single node system, i.e., a single storage server. As the number of storage servers increases, the network contention and data orchestration challenges exponentially increase.  Furthermore, each server performs more remote accesses,  increasing the contention even more. This leads to an exponential growth in latency with the increase in the number of storage servers used per application. It is recommended to contain most of the data belonging to the same application in the same storage server to minimize remove disk accesses over network.

\begin{figure}[h]
\centering 
\includegraphics[width=0.7\linewidth]{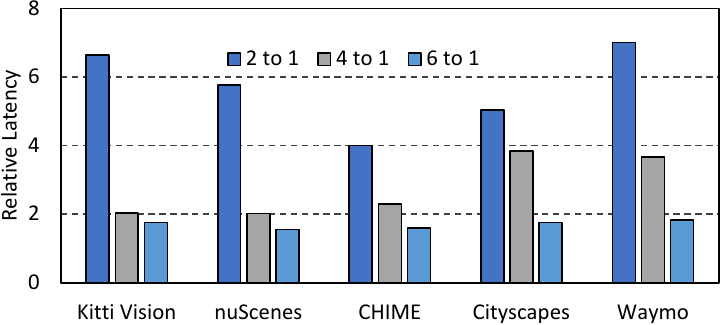}
%\vspace{-6pt}
\caption{Impact of increasing the number of CSDs in the system (the baseline system has a SSD-to-CSD ratio 8 to 1).}
\label{fig:CSDperSSD}
\end{figure}

Similarly, Fig.~\ref{fig:CSDperSSD} shows the impact of increasing the number of CSDs in a storage server. With increasing the number of CSDs per SSD (or other storage element) does show significant improvement because of increase in parallelism. However, a typical CSD is $\approx15\times$ expensive than a standard SSD and $\approx25\times$ expensive than a classical HDD. Integrating more number of CSDs into the standard storage server will significantly increase the cost of the server. Moreover, in large-scale systems where failure is common, replacing failed CSDs would further increase the cost. From our evaluation, we found an 8:1 ratio of SSD to CSD (capacity ratio) provides the best possible cost-to-acceleration benefit. 

%%%%%

% %-------------------------------------------------------------------------------

% %-------------------------------------------------------------------------------
%\vspace{-4pt}
\section{Conclusions}
\label{sec:conclusion}
%%%%%

In this paper, we have explored the critical role of storage systems in the context of continuous learning video analytics edge server, emphasizing in particular the transformative potential of Computational Storage Devices (CSDs) in this domain. Our proposal, \salt, highlights the need for a paradigm shift in storage architecture to accommodate the dynamic and computationally intensive nature of modern ML applications. By reducing unnecessary data movement and enabling near-data processing, CSDs enhance the efficiency, performance, and sustainability of storage servers. \salt, with its intelligent data orchestration and acceleration, can provide up to $6.18\times$ latency and $6.13\times$ data movement reduction, compared to classical systems. This is particularly evident in continuous learning scenarios prevalent in applications such as autonomous driving and urban mobility, where the volume and complexity of data are immense. 

Looking ahead, the role of storage systems is poised to become even more pivotal as ML applications continue to evolve and generate larger, more complex datasets. The ongoing innovation in the design and optimization of CSDs will be crucial in meeting these challenges. In summary, our research presents a comprehensive vision for the future of storage systems in ML, where computational storage devices play a key role in advancing the, performance, efficiency, and capabilities of storage servers, thereby contributing significantly to the broader field of ML. 

%%%%%

% %-------------------------------------------------------------------------------

%%%%%%% -- PAPER CONTENT ENDS -- %%%%%%%%
%%%%%%%%%%%%%%%%%%%%%%%%%%%%%%%%%%%%%%%%%%%%%%%%%%%

%%%%%%%%%%%%%%%% CONTENT ENDS HERE %%%%%%%%%%%%%%%%

%\bibliographystyle{plainnat}
\bibliographystyle{iclr2025_conference}
\bibliography{refs}

%%%%%%%%%%%%%%%% CONTENT ENDS HERE %%%%%%%%%%%%%%%%

\end{document}